\documentclass[aps,article,twocolumn]{revtex4}
\usepackage{amsmath,amssymb,amsfonts,graphicx,hyperref}
\newcommand{\beq}{\begin{equation}}
\newcommand{\eeq}{\end{equation}}
\newcommand{\bea}{\begin{eqnarray}}
\newcommand{\eea}{\end{eqnarray}}

\begin{document}

\title{Some remarks on Relativistic Diffusion and the Spectral Dimension Criterion}
\author{C. R. Muniz}
\email{celio.muniz@uece.br}\affiliation{Grupo de F\'isica Te\'orica (GFT), Universidade Estadual do Cear\'a, Faculdade de Educa\c c\~ao, Ci\^encias e Letras de Iguatu, Iguatu, Cear\'a, Brazil.}
\author{M. S. Cunha }
\email{marcony.cunha@uece.br}\affiliation{Grupo de F\'isica Te\'orica (GFT), Centro de Ci\^encias e Tecnologia, Universidade Estadual do Cear\'a, CEP 60714-903, Fortaleza, Cear\'a, Brazil.}

\author{R. N. Costa Filho}
\email{rai@fisica.ufc.br} \affiliation{Departamento de F\'isica,
  Universidade Federal do Cear\'a, Caixa Postal 6030, Campus do Pici,
  60455-760 Fortaleza, Cear\'a, Brazil.}

\author{V. B. Bezerra}
\email{valdir@fisica.ufpb.br}\affiliation{Departamento de F\'isica, Universidade Federal da Para\'iba,
Caixa Postal 5008,\\CEP 58059-970, Jo\~ao Pessoa-PB, Brazil.}

\begin{abstract}

The spectral dimension $d_s$ for high energies is calculated using the Relativistic Schr\"{o}dinger Equation Analytically Continued (RSEAC) instead of the so-called Telegraph's Equation (TE), in both ultraviolet (UV) and infrared (IR) regimens. Regarding the TE, the recent literature presents difficulties related to its stochastic derivation and interpretation, advocating the use of the RSEAC to properly describe the relativistic diffusion phenomena. Taking into account that the Lorentz symmetry is broken in UV regime at Lifshitz point, we show that there exists a degeneracy in very high energies, meaning that both the RSEAC and the TE correctly describe the diffusion processes at these energy scales, at least under the spectral dimension criterion. In fact, both the equations yield the same result, namely, $d_s = 2$, a dimensional reduction that is compatible with several theories of quantum gravity. This result is reached even when one takes into account a cosmological model - the De Sitter one - for a flat Universe. On the other hand, in the IR regimen, such degeneracy is lifted in favor of the approach via TE, due to the fact that only this equation provides the correct value for $d_s$, which is equal to the actual number of spacetime dimensions, {\it i.e.}, $d_s = 4$, while RSEAC furnishes $d_s=3$, so that a diffusing particle described by this latter experiences a three-dimensional spacetime.

\vspace{0.75cm}
\end{abstract}


\maketitle


The Classical Diffusion Equation (CDE) has been studied in a broad context ranging from problems in thermal, electrical, and nuclear engineering, passing by biological ones such as the diffusion of nutrients in the ocean \cite{Okubo}, to social and economic issues like the diffusion of new products on the market \cite{Mahajan}. In nuclear physics, the neutron flux in a nuclear reactor can be obtained from the CDE solution involving both spatially and temporally variable parameters of diffusion, which also includes terms of absorption and production of neutrons (sinks and sources). This solution is quite well known and provides the balance of the  particles average number in an infinitesimal volume of the material, which encloses the reactor core for non-relativistic (slow, thermal) neutrons \cite{Stacey}. However, when one goes to systems with relativistic velocities or high energies as in astrophysical and cosmological scenarios, there is still no agreement about the form of the diffusion equation itself, as well as on the physical interpretation of their solutions based on microscopic stochastic collisions \cite{Herrmann}.

The first detailed studies on relativistic diffusion processes were carried out independently by Rudberg in 1957 \cite{Rudberg}, and  Schay in 1961 \cite{Schay}. But it was only in the middle 80's that the relativistic diffusion drawn more attention, when one has considered the possibility of extending microscopic and stochastic mechanisms to the framework of the Special Relativity. According to those authors, any relativistic generalization of CDE with constant coefficients should be at least of second order in time. Besides this term, the so-called Telegraph's Equation (TE) preserves all the CDE terms. This commonly used relativistic diffusion equation can be formally obtained by simply replacing in CDE the Laplacian operator by the d'Alambertian one, with the temporal variable becoming the proper time. A stochastic, {\it i.e.}, random walk based derivation of TE shows that it is given by
\begin{equation}\label{00}
\beta\frac{\partial^2\rho({\bf x},{\bf x'}; \tau,\tau')}{\partial\tau^2}+\frac{\partial\rho({\bf x},{\bf x'}; \tau,\tau')}{\partial\tau}=D\Delta\rho({\bf x},{\bf x'}; \tau,\tau'),
\end{equation}
where $D$ is the diffusion coefficient dependent on the propagation speed, which cannot be arbitrarily large as in CDE; $\triangle$ is the Laplacian operator, which in curved spaces is given by $\Delta\equiv|g|^{-1/2}\partial_i(|g|^{1/2}g^{ij}\partial_j)$ and corresponds to the Laplace-Beltrami operator, and  $\beta>0$ is the relaxation time parameter, that also measures the correlation in microscopic motion, namely, how each particle continues to move itself in the same direction as previously. Such quantity introduces memory leading to a non-Markovian process \cite{Dunkel} arising from the random structure of TE.

A recently published paper \cite{boris} points out difficulties related to the stochastic derivation and interpretation of TE, advocating the use of another equation to properly describe relativistic diffusion phenomena. This equation is built from the relativistic kinetic energy operator for a free massive particle described in quantum mechanics, given by $K=\sqrt{p^2+m^2}-m$, identifying $p^2$ with $-\triangle$ and $K$ with $i\partial/\partial t$ ($c=\hbar=1$), and after this, doing a Wick rotation, $t\rightarrow-i\tau$. The obtained equation is called Relativistic Schr\"{o}dinger Equation Analytically Continuated (RSEAC), and it seems to describe diffusive processes in high energies without ambiguities, better than the TE, since the stochastic process associated to the former does not present sharp (singular) propagation wavefronts which arises in the latter. Besides this, the RESEAC also describes a non-Markovian diffusion process.

In very high energies (possibly at Planck scale or beyond), some fundamental theories require the Lorentz symmetry breaking, and a natural question that arises is about the correct diffusion equation at these scales, since that we would have a type of non-relativistic behavior again. The Ho\v{r}ava-Lifshitz theory, for example, is a recent attempt to quantize gravity \cite{horava1,horava2,horava3}, whose renormalizability via power-counting is warranted through anisotropy between space and time directions existing at ultraviolet (UV) scales. Such anisotropy is implemented by means of scaling differently time and space, in the form $t\rightarrow a^zt$ and $x^{i}\rightarrow ax^{i}$, where $a$ is a scale factor and $z$ is a dynamical critical exponent that goes to the unity at large distances, restoring General Relativity at infrared (IR) scale. In a $d$-dimensional space, the renormalizability of the theory is warranted for $z=d$, at least.

In this paper, we will compare the two relativistic diffusion mechanisms described by RSEAC and TE, by calculating the spectral dimension associated to them, in both UV and IR regimens. This quantity can be interpreted as an effective spacetime dimension experienced by a diffusing particle. In one of his seminal papers, Ho\v{r}ava \cite{horava3} used the TE to calculate the spectral dimension of the Universe, finding
\begin{equation}\label{01}
d_s=1+\frac{d}{z},
\end{equation}
so that at UV scale, $d_s=2$, and at IR one, $d_s=4$. This expression does not depend on any temporal parameter, although it is possible obtaining a continuous interpolation between these two limits through the diffusion time \cite{Brito}. Then, we will employ the Ho\v{r}ava's procedure to obtain the spectral dimension in these limits, now considering in this analysis the RSEAC. The calculations will also be performed taking into account the cosmological evolution for a flat universe at UV scale, using the model which is possibly associated with its inflationary phase -  the De Sitter one, which is also compatible with the Ho\v{r}ava-Lifshitz gravity as shown in \cite {masato}.



Let us calculate the spectral dimension of the spacetime using a diffusion law different from that one employed in Ho\v{r}ava's paper \cite{horava3} and generalized in \cite{visser2}. The principle behind this is that at very short distances the spacetime behaves as a microscopically chaotic and discrete system, with its evolution obeying purely stochastic laws. Then, instead of TE we assume that, in the continuum limit, the mathematical law governing this kind of process is the RSEAC, which is given by \cite{boris}
\begin{equation} \label{1}
\frac{\partial\rho({\bf x},{\bf x'};\sigma)}{\partial\sigma}=(m-\sqrt{m^2-\triangle})\rho({\bf x},{\bf x'};\sigma),
\end{equation}
where $\sigma$ is an external temporal parameter and $m$ is related to the diffusion coefficient. When $m=0$, we have the fractional Schr\"{o}dinger equation analytically continued \cite{laskin}. The isotropic point-source solution of Eq. (\ref{1}) is given by
\begin{align} \label{2}
P(\sigma)&=\rho({\bf x},{\bf x'};\sigma)|_{{\bf x}={\bf x'}}=\nonumber\\
&C_d\int_0^{\infty}k^{d-1}\exp{[\sigma(m-\sqrt{m^2+k^2})]}dk,
\end{align}
which is integrated in the momenta space and $C_d$ is a constant which depends on the spatial topological dimension, $d$. The Lorentz symmetry breaking predicted in the Ho\v{r}ava-Lifshitz theory will be considered by introducing the dynamical critical exponent, $z$, in Eq. (\ref{2}) in such way that $k^2\rightarrow k^{2z}$.

Next, we calculate the spectral dimension, $d_s$, by means of its definition
\begin{equation} \label{3}
d_s=-2\frac{d\log{P(\sigma)}}{d\log{\sigma}}.
\end{equation}
For our purpose, it is convenient to rewrite the spectral dimension (\ref{3}) as
\begin{equation}\label{4}
d_s=-\frac{2\sigma}{P(\sigma)}\frac{dP(\sigma)}{d\sigma}.
\end{equation}
Then, substituting Eq. (\ref{2}) into Eq. (\ref{4}), we get
\begin{equation}\label{4.a}
d_s=-\frac{2 \sigma \int_0^{\infty } k^{d-1}g(k)\exp[{\sigma g(k)}]dk}{\int_0^{\infty } k^{d-1} \exp[{\sigma g(k)}]dk},
\end{equation}
where $g(k)=\left(m-\sqrt{k^{2 z}+m^2}\right)$.

A numerical analysis of Eq. (\ref{4.a}) shows that, in UV limit, when $z=d$ and $\sigma\rightarrow0$, we have $d_s=2$ for any value of $m$. This is in total agreement with the result found in \cite{horava3} as well as with several other quantum gravity theories (\cite{carlip} and references therein).

Now, we modify the diffusion equation by introducing the Laplace-Beltrami operator, in order to consider the Friedman-Robertson-Walker (FRW) metric, which is described by
\begin{equation}\label{5}
ds^2=-dt^2+\frac{a^2(t)}{(1+\kappa r^2/4)^2}\delta_{ij}dx^idx^j,
\end{equation}
in isotropic coordinates \cite{mukhanov}. Taking into account the spacetime anisotropy via $k^2\rightarrow k^{2z}$, the solution of Eq. (\ref{1}) can be written as
\begin{equation} \label{6}
P(\sigma)=C_d\int_0^{\infty}k^{d-1}\exp{[\sigma(m-\sqrt{m^2+a^{-2}(\sigma)k^{2z}})]}dk,
\end{equation}
in which we do $t\equiv\sigma$ and take a zero-curvature Universe, $\kappa=0$. Now let us consider a FRW cosmological model compatible with the Horava-Lifshitz gravity - the De Sitter model \cite{masato}, which probably prevailed in the primordial universe, during the inflationary phase, therefore in very high energies. In this model the scale factor varies as
\begin{equation}
a(\sigma)=a_0\exp{(H_0\sigma)}.
\end{equation}
Plugging the above equation into Eq. (\ref{6}) and this into Eq. (\ref{4}), the numerical analysis  still yields $d_s=2$, where $\sigma\rightarrow0$, $z=d$ (UV regimen), for every $m$, $a_0$, and $H_0$ (all greater than zero). In particular, if we make the same analysis for TE, from this cosmological point of view, we obtain an identical result.

On the other hand, without taking into account any cosmological model, in the IR limit ($d=3$, $z=1$, and $\sigma\rightarrow\infty$) we obtain $d_s=3$. This is an unexpected result, since that the correct spectral dimension value should be $d_s=4$, which is the actual number of spacetime topological dimensions, as it was found in the Ho\v{r}ava's paper \cite{horava3}. Thus, the shortcoming of RESEAC consists in describing a particle that experiences a three-dimensional spacetime at IR scale.


In summary, we have assumed that the RSEAC is the correct law that describes relativistic processes of diffusion, as it is claimed in the recent literature \cite{boris}. Then, we have calculated the spectral dimension of the Universe using this equation, instead of that one usually employed in relativistic diffusion models - the TE. By means of a numerical analysis, we found that $d_s=2$ at UV scale, {\it i.e.}, when the diffusion time ($\sigma$) tends to zero and the anisotropic parameter ($z$) is numerically equal to the space topological dimension. With this, we can assert that there exists a degeneracy in very high energies, since that the two relativistic diffusion models provide the same spectral dimension, at least in those theories in which the Lorentz symmetry is broken at UV scale, as it is in the Ho\v{r}ava-Lifshitz gravity.

The calculations were also performed taking into account the De Sitter cosmological model, by considering the corresponding FRW metric in RSEAC, and they have given the same result. The mentioned degeneracy was reinforced when we did the same procedure, {\it mutatis mutandis}, by introducing De Sitter model in TE with anisotropic scaling, again obtaining $d_s=2$. Then the Horava's work \cite{horava3} was extended by adding this cosmological feature.

At IR scale, namely, when $\sigma\rightarrow\infty$, $d=3$, and $z=1$, where we no longer consider cosmological models, our approach gave the result $d_s=3$, which is unsatisfactory, since the expected value should be equal to the actual number of topological dimensions of the physical spacetime. In fact, such coincidence ($d_s=4$) happens when one uses the TE in the context of different quantum gravity approaches \cite{carlip}. Thus, the trouble with RESEAC in IR region is in the fact that this equation does not furnish the correct spacetime dimension under the perspective of a particle diffusing in the corespondent manifold, despite the RESEAC present a virtue that the TE does not own: is free of singularities with respect to the wavefronts that it describes. In fact, both the diffusion equations have qualities and defects, which probably claims by a law that generalizes them.

Finally, we state that the analysis by means of the spectral dimension presented in here offers a reasonable criterion for choosing among different mathematical models, which describe relativistic diffusion phenomena at distinct energy scales.


V.B.Bezerra and R.N. Costa Filho would like to thank CNPq for partial financial support.

\end{document}